\documentstyle[12pt,titlepage]{article} 
\begin{document}
\title{Non-equilibrium Thermodynamics and Hydrodynamic Fluctuations}
\author{J. M. Rub\'{\i}$^a$ and P. Mazur $^b$ \\
$^a$ {\it Departament de F\'{\i}sica Fonamental, Facultat de F\'{\i}sica,} \\
{\it Universitat de Barcelona, Diagonal, 647 08028 Barcelona, Spain} \\ 
$^b$ {\it Instituut-Lorentz, University of Leiden, P.O. Box 9506,} \\
{\it 2300 RA Leiden, The Netherlands}}
\date{ }
\maketitle
\newpage

\begin{abstract}
The reformulation of nonequilibirum thermodynamics, to include the 
treatment of thermodynamic fluctuations, is applied to the hydrodynamic 
fluctuations of a simple fluid. It is shown that the nonequilibrium thermodynamic 
scheme leads to the explicit form of the Fokker-Planck equation which 
describes the time behaviour of the probability distribution function of 
these hydrodynamic fluctuations as well as the irreversible processes which 
are connected with this behaviour.
\end{abstract}              

\section{Introduction}
In a recent paper [1] one of the authors (P.M.) showed that the scheme and 
formalism of nonequilibirium thermodynamics can be reformulated in such a way 
that the equations describing the dynamics of thermodynamic 
fluctuations are obtained in a simple manner. In particular one is thus led directly to a theory of 
fluctuations as Gaussian Markov processes obeying a multivariate Fokker-Planck 
equation [2,3]. In this reformulation of thermodynamics of irreversible 
processes fluctuations were considered as internal degrees of freedom of a 
thermodynamic system. Moreover Gibbs entropy postulate was used to define 
the system's entropy as a functional of the probability density in internal 
coordinate space.

In this paper we apply the formalism developed in [1] to the case of  
hydrodynamic fluctuations in a simple fluid. It is our aim to show that 
nonequilibrium thermodynamics yields indeed the multivariate Fokker-Planck 
equation describing the behaviour in time of hydrodynamic fluctuations and 
thus implicitely also the Langevin-like equations of fluctuating hydrodynamics 
[4].

In Section 2 we introduce the hydrodynamic fluctuations as fluctuation of the 
Fourier transformed (in space) fields of mass, density, temperature and the 
three components of fluid velocity. We also introduce in this section the 
probability density function of these fluctuations as well as the system's 
entropy as a functional of this distribution function. This functional for the 
entropy which is an expression for Gibb's entropy postulate contains also the 
equilibrium distribution function. We give an explicite calculation in 
Gaussian approximation of the latter in section 3 (and the appendix). In 
Section 4 we calculate the entropy production when the system is not in 
equilibrium. We obtain for this quantity an expression which is, in analogy to 
its form in macroscopic nonequilibrium thermodynamic [5], a sum of 
products of dissipative fluxes, which are components of the probability flux, 
and conjugate thermodynamic forces, which are derivatives of probability 
densities. As usual in thermodynamics of irreversible processes linear 
relations containing phenomenological coefficients are established between 
fluxes and forces. When theses are introduced into the continuity equation for 
the time dependent probability density the differential equation, the Fokker-
Planck equation, governing the dynamics of the hydrodynamic fluctuations as 
Markov processes, is obtained. We then determine in section 5 the 
phenomenological, or Onsager coefficients, introduced formally in section 4 
by deriving mean regression laws from the Fokker-Planck equation which was 
found, and comparing these with the macroscopic equations for the 
corresponding irreversible processes.

We conclude with a few remarks, made also previously regarding the 
applicability of the formalism used. 

\section{Hydrodynamic fluctuations as internal degrees of freedom of a 
thermodynamic system}
We consider a simple fluid enclosed in a cubic container of volume $V$, and 
having constant total mass $M$ and total energy $E$. The macroscopic 
thermodynamic internal state at time $t$ and position $\vec{r}$, of the 
system may be characterized by 5 fields $A_i (\vec{r},t)$ corresponding 
to the 5 conserved quantities $\rho (\vec{r},t)$, $e(\vec{r},t)$, 
$\rho \vec{v} (\vec{r},t)$ where $\rho$ is the mass 
density, $e$ the energy per unit of volume and $\textstyle{\vec{v}}$ the local velocity.

For our purpose it is more convenient to use an alternative and equivalent set of 
fields which we choose as follows
\begin{eqnarray}
A_1 (\vec{r},t) & =& \rho(\vec{r},t) \nonumber \\
A_2 (\vec{r},t) & =& T (\vec{r},t) \nonumber \\
A_i (\vec{r},t) &=& v_\alpha (\vec{r},t), (i=3,4,5 ; \quad \alpha = 
x,y,z), 
\end{eqnarray}
where $T(\vec{r},t)$ is the temperature field and where the index $\alpha$ denotes 
the 3 cartesian components of $\vec{v}$. In the equilibrium state the fields $A_i$ 
have uniform constant values $A_i^0$
\begin{equation}
A_1^0 =\rho_0 = \frac MV ; \quad A_2^0 =T_0 ;\quad A_i =0, \qquad i=3,4,5
\end{equation}

It is around these values that the fields $A_i(\vec{r},t)$ will fluctuate due 
to 
thermal agitation. To deal with these fluctuations a statistical description 
is needed, a statistical description which can best be effectuated in term of the 
Fourier components of the fields (2.1). To introduce these we choose for 
simplicity's sake periodic boundary conditions and write
\begin{equation}
A_i(\vec{r},t) = V^{-1} \sum_{\vec{k}} e^{i\vec{k}\cdot\vec{r}} A_i (\vec{k}, t)
\end{equation}                                                         
with
\begin{equation}
A_i(\vec{k} ,t) =\int_V d\vec{r} e^{-i\vec{k}\cdot\vec{r}} A_i (\vec{r}, t)
\end{equation}
\noindent
The wave vector $\vec{k}$ in eq. (2.3) and (2.4) will assume a denumerable set of 
values.

We shall now introduce the probability density $P( \{ A_i (\vec{k}) \} ,t)$ that 
the system is at time $t$ in a state defined by the set of values
 $\{ A_i (\vec{k}) \} = \underline{A}$. To normalize this probability
density 
we introduce a cut-off wave number $k_{max}$ such that no state occurs for 
which $|\vec{k}| \geq k_{max}$. The wave number $k_{max}$ in fact defines a 
length scale which is much larger than the molecular scale but small enough 
for a hydrodynamic and thermodynamic description to hold. With the above 
considerations in mind we now normalize the probability density as follows
\begin{equation}
\int_{|\vec{k}|\leq k_{max}}  P(\underline{A}, t) d\underline{A} = 1
\end{equation}

In the limit as $t\rightarrow\infty$ the distribution function 
$P(\underline{A},t)$ becomes the 
equilibrium distribution function $P_0(\underline{A})$ for which the state variables 
have mean values $\langle A_i (\vec{k})\rangle$
\begin{equation}
\langle A_i (\vec{k})\rangle = \int A_i (\vec{k}) P_0(\underline{A}) 
d\underline{A}
\end{equation}

Substituting eq. (2.4) into eq. (2.6) we have
\begin{equation}
\langle A_i (\vec{k})\rangle = V \langle A_i(\vec{r})\rangle \delta_{\vec{k},0}
\end{equation}
where we have used the fact that $\langle A_i(\vec{r})\rangle$ is independent 
of $\vec{r}$ and that
\begin{equation}
\delta_{\vec{k},0} = V^{-1} \int d\vec{r} e^{i\vec{k}\cdot\vec{r}}
\end{equation}
Since the equilibrium averages of $A_i(\vec{r})$ should be equal to the 
values of the equilibrium state given in eq. (2.2) it follows that
\begin{equation}
\langle A_1 (\vec{k})\rangle = M\delta_{\vec{k},0}; \quad 
\langle A_2 (\vec{k})\rangle = V T_0\delta_{\vec{k},0}; \quad
\langle A_i (\vec{k})\rangle = 0, \qquad  i=3,4,5.
\end{equation}

We now define fluctuations
\begin{equation}
\alpha_i (\vec{k}) = A_i(\vec{k}) -\langle A_i (\vec{k})\rangle
\end{equation}
These obey the distribution functions $P(\underline{\alpha},t)$ and $P_0 
(\underline{\alpha},t)$ 
introduced above. We also note at this point that the fluctuations $\alpha_i 
(\vec{k})$ vanish for $\vec{k} =0$: while this is true rigorously for 
$\alpha_1$ as a consequence of mass conservation it is also true to linear 
order in fluctuations (that is to the order relevant for our developments) as 
a consequence of energy and momentum conservation.

We may now consider the fluctuations $\alpha_i(\vec{k})$ as internal degrees of 
freedom of the fluid in the sense of nonequilibrium thermodynamics [6] as was done 
in the more formal preceding paper [1]. In this perspective in which the fluid is 
considered as a many component system with mass densities $M P(\underline{ 
\alpha},t)$, 
its total entropy $S$ obeys a Gibb's equation
\begin{equation}
\delta S = - M \int d\underline{ \alpha} \frac{\mu (\underline{\alpha})}{T_0} \delta P + 
\frac{1}{T_0} 
\delta E - \frac{\mu_0}{T_0} \delta M 
\end{equation}

\noindent
where $\mu (\underline{\alpha} )$ is the chemical potential of the component with mass 
density $MP(\underline{\alpha} ,t)$ and $\mu_0$ the chemical potential per unit of 
mass of the fluid system at equilibrium.

The entropy $S(t)$ can moreover be expressed as a functional of 
$P(\underline{\alpha} ,t)$ 
according to Gibb's entropy postulate
\begin{equation}
S = -k_B \int P(\underline{ \alpha} ,t)\ln\frac{P(\underline{\alpha} 
,t)}{P_0(\underline{\alpha} )} d\underline{\alpha} + 
S_0 (E,M)
\end{equation}

\noindent
where $S_0$ is the entropy at equilibrium and $k_B$ Boltzmann's constant.
This expression yields for the differential of $S$
\begin{equation}
\delta S= -k_B \int d\underline{\alpha}\ln \frac{P(\underline{ \alpha} 
,t)}{P_0(\underline{\alpha} )} \delta P +
\frac{1}{T_0}\delta E -\frac{\mu_0}{T_0} \delta M
\end{equation}

\vspace*{.3cm}\noindent
Comparing eq. (2.11) and eq. (2.13) and using eq. (2.5) we find for $\mu 
(\underline{\alpha})$
\begin{equation}
\mu (\underline{\alpha}, t) = 
\frac{k_BT_0}{M}\ln\frac{P(\underline{\alpha},t)}{P_0(\underline{\alpha})}+ \mu_0
\end{equation}

\vspace*{.3cm}\noindent
The previous formulae in particular eqs. (2.11) and (2.13) on which our further 
analysis will be based indicate the essential role played by the equilibrium 
distribution function $P_0 (\underline{\alpha})$. It is for this reason that we shall now 
first establish the explicit form of this function for the variables 
$\alpha_i(\vec{k})$ introduced above.

\setcounter{equation}{0}
\section{The equilibrium distribution function of fluctuations}

To discuss the equilibrium distribution function $P_0 (\underline{\alpha} )$ we use the 
connexion between the probability of a given state $\underline{\alpha}$ and the value of 
its entropy $S(\underline{\alpha} )$
\begin{equation}
P_0 (\underline{\alpha} ) = \mbox{const}\times  e^{S(\underline{\alpha} )/k_B}
\end{equation}
or 
\begin{equation}
P_0 (\underline{\alpha} ) = P_0(0) e^{\Delta S(\underline{\alpha})/k_B}
\end{equation}
with
\begin{equation}
\Delta S (\underline{\alpha} ) = S(\underline{\alpha}) - S(0)
\end{equation}
where $S(0)$ refers to the entropy of the equilibrium state.

We will therefore now calculate $\Delta S(\underline{\alpha})$ to second order in the 
fluctuating variables which we have introduced in section 2. We first observe 
that the entropy of our fluid may be written as the volume integral of an 
entropy density per unit of volume $s(\vec{r})$
\begin{equation}
\Delta S = \int_V \Delta s (\vec{r}) d\vec{r}
\end{equation}

At each point $\vec{r}$ the entropy density $s(\vec{r})$ and its 
fluctuation $\Delta$ will be a function of the internal energy density 
$u(\vec{r})$ and the mass density $\rho (\vec{r})$.

If we then calculate $\Delta S$ to second order in $\Delta u$ and $\Delta \rho$ 
which is sufficient within the context of a linear fluctuation theory we have
$$\Delta S = \int d\vec{r}\left\{ \left( \frac{\partial s}{\partial u}
\right)_\rho \Delta u + 
\left( \frac{\partial s}{\partial \rho }\right)_u \Delta \rho + + \frac 12  
\left( \frac{\partial^2 s}{\partial u^2}\right) (\Delta u )^2 +\right.$$
\begin{equation}
\left. \frac{\partial^2 s}{\partial u\partial \rho} \Delta u\Delta \rho + \frac 12 
\frac{\partial^2 s}{\partial \rho^2} (\Delta \rho )^2 \right\}
\end{equation}

In the appendix, we show that this expression can be reduced to a quadratic form in 
the fluctuations $\Delta \rho(\vec{r})$, $\Delta T(\vec{r})$ and 
$\vec{v}(\vec{r})$ or alternatively 
$\Delta \rho (\vec{k})$, $\Delta T (\vec{k})$ and $\vec{v}(\vec{k})$
\begin{eqnarray}
\Delta S & = - \frac{1}{2T_0} \int d\vec{r} \{ \frac{1}{\rho_0} 
(\frac{\partial p}{\partial \rho})_{T_0} (\Delta \rho )^2 + \frac{\rho_0 c_v}{T_0} 
(\Delta T)^2 + \rho_0v^2 \} \nonumber \\
& \nonumber \\
& = - \frac{1}{2T_0V} \sum_{\vec{k}} \{ \rho_0^{-1}\beta \Delta \rho (\vec{k}) 
\Delta \rho (-\vec{k}) +  \nonumber \\
& \nonumber \\
&\frac{\rho_0 c_v}{T_0} \Delta T (\vec{k}) \Delta T (-\vec{k})
 + \rho_0 \vec{v}(\vec{k}) \cdot \vec{v}(-\vec{k}) \}
\end{eqnarray}
where $p$ is the pressure and $c_v = \rho_0^{-1} \partial u/\partial T$ the 
specific heat per unit of mass; $\beta =(\partial P/\partial \rho )_T$ denotes the 
isothermal compressibility. The fluctuating variables in the sum over $\vec{k}$ 
correspond to the five variables $\alpha_i(\vec{k})$, $i= 1,2,\cdots ,5$ 
introduced in section 2.

We shall redefine in a minor way three of these variables, namely those for $i=3,4,5$ 
and rewrite the scalar product $\vec{v} (\vec{k})\cdot \vec{v} (-\vec{k})$ 
as follows
\begin{equation}
\vec{v} (\vec{k})\cdot \vec{v} (-\vec{k}) =
v (\vec{k}) v (-\vec{k}) + \vec{v}^T (\vec{k})\cdot \vec{v}^T (-\vec{k})
\end{equation}
where the scalar function $v (\vec{k})$ and the transverse vector 
$\vec{v}^T (\vec{k})$ are given by
\begin{equation}
v (\vec{k}) = \hat{k}\cdot\vec{v} (\vec{k})
\end{equation}
\begin{equation}
\vec{v}^T (\vec{k}) = (1- \hat{k}\hat{k})\cdot\vec{v}(\vec{k})
\end{equation}
with $\hat{k}$ the unit vector in the direction of $\vec{k}$, $\hat{k} = 
\vec{k}/k$. Defining now the modified variables
\begin{equation}
\alpha_3 (\vec{k}) = v (\vec{k})
\end{equation}

\begin{equation}
\alpha_i (\vec{k}) = v_\gamma^T (\vec{k}), \qquad (i= 4,5; \gamma = 1,2),
\end{equation}
with $\gamma$ denoting the transverse orthogonal components of 
$\vec{v}^T(\vec{k})$, $\Delta S$ can be written as
\begin{equation}
\Delta S = - \frac 12 \sum_{\vec{k},ij} g_{ij} (\vec{k}) \alpha_i(\vec{k}) 
\alpha_j (-\vec{k})
\end{equation}
The real matrix $g_{ij}(\vec{k})$ in this quadratic form is diagonal, $g_{ij} = 
g_{ii} \delta_{ij}$ and has elements
\begin{equation}
\left. \begin{array}{ll}
g_{11} =(V T_0 \rho_0)^{-1}\beta ; & g_{22}= (V T_0^2)^{-1} \rho_0 c_v \\
& \\
g_{ii} = (V T_0)^{-1} \rho_0; & i= 3,4,5 
\end{array}
\right\}
\end{equation}

Formulae (3.2), (3.12) and (3.13) completely determine the equilibrium 
distribution function $P_0(\underline{\alpha} )$ and yield for variables $X_i (\vec{k})$, 
conjugate to $\alpha_i (\vec{k})$ according to
\begin{equation}
X_i (\vec{k}) \equiv -k_B \frac{\partial\ln P_0}{\partial\alpha_i (\vec{k})}=
\sum_j g_{ij} \alpha_j (\vec{k}),
\end{equation}
the relations
\begin{equation}
\left. \begin{array}{ll}
X_1 (\vec{k}) = (T_0 V \rho_0)^{-1} \beta\Delta \rho (\vec{k}), &
X_2 = (V T^2_0)^{-1}\rho_0 c_v \Delta T(\vec{k}) \\
& \\
X_i(\vec{k}) = (T_0V)^{-1} \rho_0 \vec{v}(\vec{k}), & i=3,4,5 
\end{array}
\right\}
\end{equation}

We shall now evaluate in the next section the entropy produced in the system 
when the hydrodynamic fluctuations do not have their equilibrium distribution 
and derive, following the rules of non-equilibrium thermodynamics, the 
Fokker-Planck equation describing the dynamics of these fluctuations.

\setcounter{equation}{0}
\section{Entropy production and the Fokker-Planck equation for hydrodynamic 
fluctuations}

Since the system considered is a closed one with constant total mass and 
energy, its rate of change of entropy with respect to time is according to 
equations (2.11) and (2.13), or equivalently (2.12)
\begin{equation}
\frac{d S}{d t} = - k_B \int d\underline{\alpha} \ln \frac{P(\underline{\alpha} ,t)}{P_0
(\underline{\alpha} )} 
\frac{\partial P(\underline{\alpha} , t)}{\partial t}
\end{equation}

We shall now make use of the continuity equation obeyed by the probability 
density $P(\underline{\alpha} ,t)$
\begin{equation}
\frac{\partial P(\underline{\alpha} ,t)}{\partial t} = - \sum_{\vec{k},i} 
\frac{\partial }{\partial\alpha_i (\vec{k})}
J_{\vec{k},i} P(\underline{\alpha} ,t)
\end{equation}
where $J_{\vec{k},i}$ can be interpreted as a current per unit of phase space 
density along the $\vec{k}, 
i$ axis of the phase space spanned by the complex state variables 
$\alpha_{\vec{k},i}$. Equation (4.2) represents as it were, the local form 
 in $\underline{\alpha}$-space of probability conservation. Notice that, since $A_i(\vec{r})$
is a real quantity, the state variables $\alpha_i(\vec{k})$ obey 
the relation $\alpha^*_i (\vec{k})= \alpha_i (-\vec{k})$. 
Consequently, volume elements in the phase-space will be real and positive and 
since probabilities are real and positive by definition so will be the probability 
density $P(\underline{\alpha} )$ as it should. Equation (4.1) then implies that the 
currents $J_{k,i}$ have the property $J^*_{\vec{k},i} = J_{-\vec{k},i}$.

We now substitute eq. (4.2) into (4.1) and obtain, integrating by parts and 
using the fact that $P(\underline{\alpha} ,t)$ tends to zero sufficiently rapidly for 
$\alpha_i (\vec{k})$ tending to infinity, for the irreversible rate of change 
of entropy
\begin{equation}
\frac{d S}{dt} = \int P(\underline{\alpha} ,t) \sigma (\underline{\alpha} ,t) 
d\underline{\alpha} \geq 0
\end{equation}
with
\begin{equation}
\sigma (\underline{\alpha} ) = - k_B \sum_{\vec{k},i} J_{\vec{k},i} 
\frac{\partial}{\partial\alpha_i (\vec{k})} \ln \frac{P(\underline{\alpha}
,t)}{P_0(\underline{\alpha} )}
\end{equation}
the local entropy production in $\underline{\alpha}$-space. 
As usual in non-equilibrium thermodynamics the entropy production is a sum of 
products of 
currents and thermodynamic forces. In the present case the latter are the 
gradients in $\underline{\alpha}$-space of the chemical potential (2.14).

The total entropy production for the system is according to eq. (4.3) an 
average over possible fluctuating states of $\sigma (\underline{\alpha})$, the entropy 
production accompanying the evolution of a particular state. Since moreover 
the inequality (4.3) is valid for an arbitrary choice of the probability 
density $P(\underline{\alpha})$, it follows that the second law of thermodynamics holds not
only for the average of 
$\sigma (\underline{\alpha})$ but also for $\sigma (\underline{\alpha})$ itself
\begin{equation}
\sigma (\underline{\alpha}) \geq 0
\end{equation}

\vspace*{.3cm}\noindent
In equilibrium when $\sigma (\underline{\alpha})$ vanishes, both the currents and 
thermodynamics forces occurring in expression (4.4) vanish. In non-equilibrium 
and within a sufficiently wide range of physical conditions we may then suppose that these
quantities are connected by linear 
relations. In a strictly linear theory of hydrodynamic fluctuations to which 
we will restrict ourselves, the coefficients in these laws, linear in the 
thermodynamic forces, must be constant with respect to the variables $\alpha_i 
(\vec{k})$, while the various modes do not couple. The linear laws will then 
be of the form
\begin{equation}
J_{\vec{k},i} = -k_B \sum^{5}_{j=1} L_{ij} (\vec{k})
\frac{\partial}{\partial\alpha_j (\vec{k})} \ln \frac{P(\underline{\alpha},t)}{P_0 
(\underline{\alpha})}
\end{equation}
If we furthermore introduce the variables $X_j (\vec{k})$  conjugate to 
$\alpha_j(\vec{k})$ according to eq. (3.14) relations (4.6) can alternatively 
be written as

\begin{equation}
J_{\vec{k},i}= -\sum_j L_{ij} \left( X_j + k_B \frac{\partial}{\partial\alpha_j(\vec{k})}
\ln P(\underline{\alpha}) \right) 
 \end{equation}

Substituting equation (4.7) into the conservation law (4.2) the following 
differential equation is obtained
\begin{equation}
\frac{\partial P}{\partial t} = \sum_{\vec{k}} \sum_{i,j} L_{ij} (\vec{k})
\frac{\partial}{\partial\alpha_i(\vec{k})}
\left\{ PX_j (\vec{k}) + k_B \frac{\partial P}{\partial \alpha_j (\vec{k})}
\right\} 
\end{equation}
which describes the time evolution of the probability density 
$P(\underline{\alpha},t)$.

With the linear relation between $X_i(\vec{k})$ and $\alpha_j (\vec{k})$ (c.f. 
eq.(3.14)), the differential equation (4.8) can also be written in the form
\begin{equation}
\frac{\partial P}{\partial t} = \sum_{\vec{k}} \sum_{i,j} 
\left\{ M_{ij}(\vec{k}) \frac{\partial}{\partial\alpha_i(\vec{k})} \alpha_j 
(\vec{k}) P + 
k_B L^S_{ij} (\vec{k}) \frac{\partial^2 P}{\partial\alpha_i 
(\vec{k})\partial\alpha_j(\vec{k})} \right\}
\end{equation}
where
\begin{equation}
M_{ij} = \sum_l L_{il} g_{lj}, \quad L_{ij} = \sum_l M_{il} g^{-1}_{lj}
\end{equation}
and
\begin{equation}
L^S_{ij} = \frac 12 (L_{ij} + L_{ji})
\end{equation}
This equation has the standard form of a linear Fokker-Planck equation and 
describes therefore the dynamics of the hydrodynamic fluctuations as Gaussian-
Markov processes. As stated previously this result has been obtained here within the
framework of 
non-equilibrium thermodynamics and its extension to system with internal 
degrees of freedom.

\setcounter{equation}{0}
\section{Identification of the Onsager coef\-fi\-cients in the Fokker-Planck 
equation}
Note that the Fokker-Planck equation (4.9) contains the scheme of coefficients 
$M_{ij}$ or equivalently $L_{ij}$ (the matrix $g_{ij}$ connecting these has 
already been specificied above) which must still be determined to make the 
description of the dynamics of the fluctuations complete. The determination 
can be achieved by observing that fluctuations regress in the mean according 
to the macroscopic laws for irreversible processes. This experimentally 
verified bahaviour is known as Onsager's regression hypothesis [7] and enables one 
to identify, in the case of linear laws the coefficients $L_{ij}$ in terms of 
macroscopic quantities.

We establish the mean regression laws from the Fokker-Planck equation (4.9) by 
multiplying both of its members by $\alpha_i (\vec{k})$ and integrating over 
all variables $\underline{\alpha}$. We then find for the first moments of the 
variables $\alpha_i$ after partial integration the ordinary differential 
equations

\begin{eqnarray}
\frac{d\overline{\alpha_i}^{\alpha_0}(\vec{k},t)}{dt} &=& -\sum_j L_{ij}(\vec{k}) 
\overline{X_j}^{\alpha_0}(\vec{k},t) \nonumber \\ 
& =& - \sum_l M_{il} (\vec{k}) \overline{\alpha_l}^{\alpha_0} (\vec{k},t),
\end{eqnarray}
where the conditional mean $\overline{\alpha_i}^{\alpha_0}$ is given by
\begin{equation}
\overline{\alpha_i}^{\alpha_0} = \int \alpha_i (\vec{k},t) 
P(\underline{\alpha},t) d\underline{\alpha}
\end{equation}
with
\begin{equation}
P(\underline{\alpha},0)= \prod_{\vec{k},i} \delta \left( \alpha_i (\vec{k})- 
\alpha_{i0} (\vec{k})\right)
\end{equation}

The mean regression equations constitute one of the fundaments of Onsager's 
derivation [7] of reciprocal relations for the coefficients $L_{ij}$, which 
follow by applying microscopic reversibility. For the equations (5.1) 
containing complex variables $\alpha_i(\vec{k})$, defined through equations 
(2.1) these reciprocal relation are [8] for complex Onsager coefficients 
$L_{ij}$,
\begin{equation}
L_{ij} = L_{ji}^* \varepsilon_i \varepsilon_j
\end{equation}
with
\begin{equation}
\varepsilon_i = \left\{ \begin{array}{rl}
                   1 & \mbox{for}\qquad i= 1,2 \\
                  -1 & \mbox{for}\qquad i= 3,4,5
                  \end{array} \right.
\end{equation}

The coefficients $L_{ij}$ satisfy, in addition to the reciprocal relations, 
certain conditions imposed by the spatial symmetry of the system. For the 
isotropic fluid considered this implies that scalar and vectorial phenomena 
do not couple so that
\begin{equation}
L_{ij} = L_{ji} = 0 \qquad \mbox{for} \{ \begin{array}{ll}
                                  i =&  1,2,3 \\
                                  j = & 4,5     \end{array}
\end{equation}
and also that the matrix $L_{ij}$, $i,j= 4,5$ is diagonal
\begin{equation}
L_{ij} = L\delta_{ij},\qquad  i, j= 4,5
\end{equation}
To identify finally the coefficients $L_{ij}$ we list below the familiar [9] 
fully linear macroscopic hydrodynamic equations for the density, temperature 
and velocity fields, $\rho (\vec{k},t)$, $T(\vec{k},t)$ and $v(\vec{k},t)$. 
These equations are
\begin{equation}
\frac{\partial\rho (\vec{k},t)}{\partial t} = -i \rho_0 k \hat{k}\cdot 
\vec{v} (\vec{k},t)
\end{equation}
\begin{equation}
\frac{\partial T(\vec{k},t)}{\partial t} = - \frac{T_0 \left( \frac{\partial 
p}{\partial T}\right)_{\rho_0}}{\rho_0c_v} i k\hat{k} \cdot \vec{v}(\vec{k},t) - 
\frac{k^2\lambda}{\rho_0 c_v} T(\vec{k},t)
\end{equation}
\begin{eqnarray}
\frac{\partial\vec{v}(\vec{k},t)}{\partial t}=&-& \frac{i\vec{k}\beta}{\rho_0} \rho 
(\vec{k},t) 
- \frac{i\vec{k}}{\rho_0} \left( \frac{\partial p}{\partial T} 
\right)_{\rho_0} T(\vec{k}, t) \nonumber \\ 
& -& \frac{1}{\rho_0} \left\{ k^2 \eta + \left( \frac{1}{3} \eta + \eta_v \right) 
\vec{k}\vec{k} \cdot \right\} \vec{v} (\vec{k}, t)
\end{eqnarray}
In equation (5.9) $\lambda$ is the heat conductivity; $\eta$ and $\eta_v$ in 
equation (5.10) are the viscosity and volume viscosity respectively.

By multiplying moreover eq. (5.10) succesively with the unit vector 
$\hat{k}$, and the projection operator $1-\hat{k}\hat{k}$ one obtains the 
macroscopic equations for the variables $v(\vec{k})$ and $\vec{v}^T (\vec{k})$ 
introduced in section 3, cf. eqs. (3.8) and (3.9).

Comparing then the mean regression equation (5.1) with the macroscopic 
hydrodynamic equations one finds for the matrix of phenomenological 
coefficients $M_{ij}$
$${\cal M}= \left(  \begin{array}{ccccc}
              0   & 0 &  i k \rho_0 & 0 & 0 \\
&&&&\\
              0   & \frac{k^2\lambda}{\rho_0c_v} & \frac{ikT_0}{\rho_0c_v}\left(\frac{\partial P}{\partial 
T}\right)_{\rho_0}&   0 &  0 \\
&&&&\\
\frac{ik\beta}{\rho_0} &\frac{ik}{\rho_0}\left(\frac{\partial P}{\partial
T}\right)_{\rho_0}&\frac{k^2}{\rho_0} (\frac 43\eta + \eta_v) & 0 & 0  \\
&&&&\\
0 & 0 & 0 & \frac{k^2\eta}{\rho_0} & 0 \\
&&&&\\
0 & 0& 0 & 0 & \frac{k^2\eta}{\rho_0} \end{array} \right) \eqno{(5.11)}$$
and for the matrix of Onsager coefficients $L_{ij}$, using also eqs. (3.13) and 
(4.10),
$$ {\cal L} = \left( \begin{array}{ccccc}
       0 & 0 & ik VT_0 & 0 & 0 \\
&&&&\\
       0 &  VT_0^2 k^2\lambda   /\rho^2_0 c_v^2  & ik (\frac{\partial P}{\partial T})_\rho 
VT_0^2/\rho^2_0c_v & 0 & 0 \\
&&&&\\
       ik VT_0 &  ik (\frac{\partial P}{\partial T})_\rho 
VT_0^2/\rho^2_0c_v & V T_0 k^2 (\frac 43 \eta +\eta_v) / \rho^2_0 & 0&0 \\
&&&&\\
0 & 0& 0 &VT_0 k^2 \eta /\rho^2_0 &0  \\
&&&&\\
0 & 0& 0 & 0& VT_0 k^2 \eta /\rho^2_0  
\end{array} \right) \eqno{(5.12)}$$

This then concludes the specification within the context of nonequilibrium 
thermodynamics of the Onsager coefficients $L_{ij}$ for the linear fluctuating 
fluid considered and completely determines the Gaussian solution of the linear 
Fokker-Planck equation. 
Observe that the reciprocal relations (5.4) as well as the 
spatial symmetry requirements (Curie's principle) (5.6) and (5.7) are 
automatically satisfied by the coefficients found above.

The above application to hydrodynamic fluctuations of the recently developed 
extension to fluctuation phenomena [1] of the scheme of macroscopic nonequilibrium thermodynamics, illustrates once more previously stated 
characteristics of this generalized formalism which goes beyond the 
traditional macroscopic domain. It introduces indeed, while keeping the 
framework and rules of the macroscopic theory, statistical notions such as:

1. The distribution function of hydrodynamic fluctuations.

2. The definition of entropy as a functional of this distribution function.

3. The calculation of the probability distribution 
function in equilibrium. 

We also stressed before that the extension of the 
thermodynamic formalism is legitimate because it is limited to events taking 
place on a time scale which is slow compared to the time scale of 
microscopic phenomena. For the case at hand the phenomena dealt with occur at 
characteristic times which are essentially hydrodynamic relaxation times.

\vspace*{.6cm}
\setcounter{equation}{0}
\noindent
{\LARGE\bf Appendix}

\vspace*{.6cm}
In this appendix we show for completeness sake that expression (3.5) for 
$\Delta S$, the variation of entropy with respect to its equilibrium value up 
to second order, reduces to the diagonal quadratic form in $\Delta 
T(\vec{r})$ and $\Delta \rho (\vec{r})$ given in formula (3.6).

We first observe that in the integral (3.5) the linear terms give rise, due to 
mass and energy conservation to the  integral of the quadratic term $\frac 12
\rho_0 v^2(\vec{r})$:
$$\int \left\{ \frac{\partial s}{\partial u} \Delta u + \frac{\partial s}
{\partial \rho}\Delta \rho \right\} d\vec{r} = - \frac{1}{2T_0} \rho_0 \int v^2 
d\vec{v}, \eqno{(A.1)}$$
since all thermodynamic derivatives in 
(A.1) are uniform equilibrium quantities and since $\partial s/\partial u= 1/T_0$. 
Next we observe that the second order 
terms in the integral (3.5) may be written in the following form
$$\frac 12 \frac{\partial^2s}{\partial u^2}(\Delta u)^2 + \frac{\partial^2 
s}{\partial u\partial \rho} \Delta u \Delta \rho + \frac{\partial^2s}{\partial \rho^2} 
(\Delta \rho)^2 = \Delta \frac 1T \Delta u - \Delta \frac \mu T \Delta \rho 
\eqno{(A.2)}$$
as
$$\Delta \frac{1}{T} = \Delta\frac{\partial s}{\partial u} = \frac{\partial^2 
s}{\partial u^2} \Delta u + \frac{\partial^2 s}{\partial u \partial \rho} 
\Delta \rho \eqno{(A.3)}$$
and
$$- \Delta \frac{\mu}{T} = \Delta \frac{\partial s}{\partial \rho} = 
\frac{\partial^2s}{\partial \rho \partial u} \Delta u + 
\frac{\partial^2s}{\partial \rho^2} \Delta \rho . \eqno{(A.4)}$$

\vspace*{.3cm}
Expand now in expression (A.2) $\Delta u$ and $\Delta\mu /T$ to linear order 
in $\Delta T$ and $\Delta \rho$
$$\Delta u = \frac{\partial u}{\partial T} \Delta T + \frac{\partial u}{\partial 
\rho} \Delta \rho \eqno{(A.5)}$$ 

$$\Delta \frac{\mu}{T} = \frac{\partial\mu /T}{\partial T} \Delta T
+  \frac{1}{T} \frac{\partial\mu}{\partial \rho} \Delta \rho \eqno{(A.6)}$$
Here again the thermodynamic derivatives are uniform equilibrium quantities.

Finally observe that the total differential of the thermodynamic potential 
$f /T$, with $f=u-Ts$ the free energy density,
$$d \frac{f}{T} = - \frac{u}{T^2} dT + \frac{\mu}{T} d\rho \eqno{(A.7)}$$
leads to the Maxwell relation
$$- \frac{\partial\mu /T}{\partial T} = \frac{1}{T^2} \frac{\partial u}{\partial 
\rho} \eqno{(A.8)}$$

Substituting then (A.1) and (A.2) with (A.5) and (A.6) into expression (3.5) and taking into 
account the Maxwell relation (A.8) one obtains the desired result (3.6).

\end{document}